\documentclass[lineno]{JFM-FLM_Au}
\usepackage{lscape}
\usepackage[dvipsnames]{xcolor}
\usepackage{soul}

\usepackage{comment}
\usepackage{natbib}
\newcommand{\eqn}[1]{\begin{align}#1\end{align}}
\newcommand{\bs}[1]{\boldsymbol{#1}}

\def\bR{\bs{R}}

\def\bt{\bs{t}}

\def\bt{\bs{t}}

\def\bu{\bs{u}}

\newcommand{\Blaise}[1]{{\bf[{\color{magenta}B: #1}]}}
\newcommand{\Marine}[1]{{\bf[{\color{purple}M: #1}]}}

\lefttitle{M. Aulnette, M. Coutadeur, C. Bielinski, B. Delmotte and A. Lindner}
\righttitle{Journal of Fluid Mechanics}

\title{Orientation Dynamics of Rigid Fibers in a Microfluidic Burgers-like Vortex}

\author{Marine Aulnette\aff{1}, Mathis Coutadeur\aff{2}, Clément Bielinski\aff{2}, Blaise Delmotte\aff{2} \and Anke Lindner\aff{1,3}}

\affiliation{\aff{1}Laboratoire de Physique et Mécanique des Milieux Hétérogènes (PMMH), ESPCI Paris, PSL University, CNRS, Sorbonne University, and Paris Cité University, 75005 Paris, France
\aff{2}LadHyX, CNRS, Ecole Polytechnique, Institut Polytechnique de Paris, 91120 Palaiseau, France
\aff{3}Institut Universitaire de France, Paris, France}

\corresau{Anke Lindner, \email{anke.lindner@espci.fr}, Blaise Delmotte, \email{blaise.delmotte@cnrs.fr}}

\begin{document}
\maketitle

\begin{abstract}
Fiber suspensions are common in biological and environmental flows and are widely used in industrial applications. Fiber transport and orientation dynamics are affected by interactions with the surrounding fluid and strongly depend on the nature of the flow. The complexity of realistic flows, which are often heterogeneous or time-dependent, hinders a full understanding of fiber dynamics. In this study, we combine microfluidic experiments, theory and numerical simulations to investigate the orientation dynamics of rigid neutrally buoyant fibers in a well-controlled model system, a streamwise stationary vortex at moderate Reynolds number. Despite the three-dimensional nature of the flow, the orientation dynamics are remarkably simple: the fiber orientation is accurately described by Jeffery equations coupled with the Burgers-vortex model over a large range of particle Reynolds number. We show that fibers undergo uniform precession about the vortex axis driven by fluid vorticity while simultaneously aligning with the latter due to strain in the vortex core. These two motions are decoupled, with the alignment timescale determined by the local strain rate and the fiber aspect ratio. Finite particle size and inertia induce weak deviations from the base flow streamlines while leaving the orientational dynamics largely unaffected. These results establish a simple framework for understanding the behavior of elongated particles in stretched vortex flows, which constitute key building blocks of turbulence.
\end{abstract}



\section{Introduction}
Flows laden with suspended particles are ubiquitous in nature and industrial processes, particularly when the particles are non-spherical or elongated \citep{pedley1992hydrodynamic}. Such anisotropic particle suspensions arise in applications ranging from paper manufacturing and composite material processing to drag-reduction strategies, and have gained renewed attention in the context of environmental microplastic pollution, where a large fraction of debris consists of fiber-like particles \citep{lundell2011fluid,dibenedetto2025fluid,clark2023dispersion,ross2021pervasive,testa2026microfibers}.

The dynamics of elongated particles are enriched by an additional orientational degree of freedom. In canonical flows, such as simple shear, extensional or steady flows, rigid fibers are advected along streamlines while continuously rotating and reorienting in response to the local velocity gradients \citep{jeffery1922}. In more complex configurations, including mixed, unsteady and turbulent flows, the competition between strain and rotation gives rise to richer orientational dynamics \citep{voth2017anisotropic,Marchioli2026}. However, the inherently unsteady nature of turbulence continually modifies the local velocity-gradient tensor, making the underlying alignment mechanisms difficult to isolate from instantaneous observations. Statistical studies have shown that elongated particles preferentially align with the direction of fluid stretching and, consequently, with the local vorticity \citep{Ni2014,Ni2015,oehmke2021spinning,parsa2012rotation,pumir2011orientation}. For sufficiently small neutrally buoyant particles in viscous flows, these dynamics are accurately described by Jeffery  equations, which relate particle rotation to the local strain-rate and rotation-rate tensors \citep{jeffery1922}. Jeffery's framework has since been generalized to account for particle activity, flexibility, asymmetry and inertia \citep{junot2019swimming,liu2018morphological,einarsson2016tumbling,di2024influence, du2019dynamics}.

Despite extensive investigations of fiber dynamics in canonical flows such as simple shear and elongation, as well as in turbulence, comparatively little attention has been devoted to stationary flow configurations combining strong rotation and strain, such as isolated vortices. Bin Islam et al. have recently investigated numerically the transport and morphology of flexible fibers in a Burgers-like vortex, with particular emphasis on the interplay between flexibility, fiber rotation and alignment with the vortex axis \citep{islam2026dancing}. Earlier studies have examined the interaction of microswimmers with Burgers vortices, which serve as simplified models of turbulent vortex tubes \citep{Jumars2009,webster2015copepods,tanasijevic2022microswimmers}. Vortical flows have also been exploited in biological applications, for example to enhance the production of extracellular vesicles from cell spheroids \citep{thouvenot2024high}. However, despite these recent advances, a detailed understanding of the transport and orientation dynamics of passive rigid fibers in stationary stretched vortices is still lacking..

\begin{figure}[t]
\centering
\includegraphics[width=13cm]{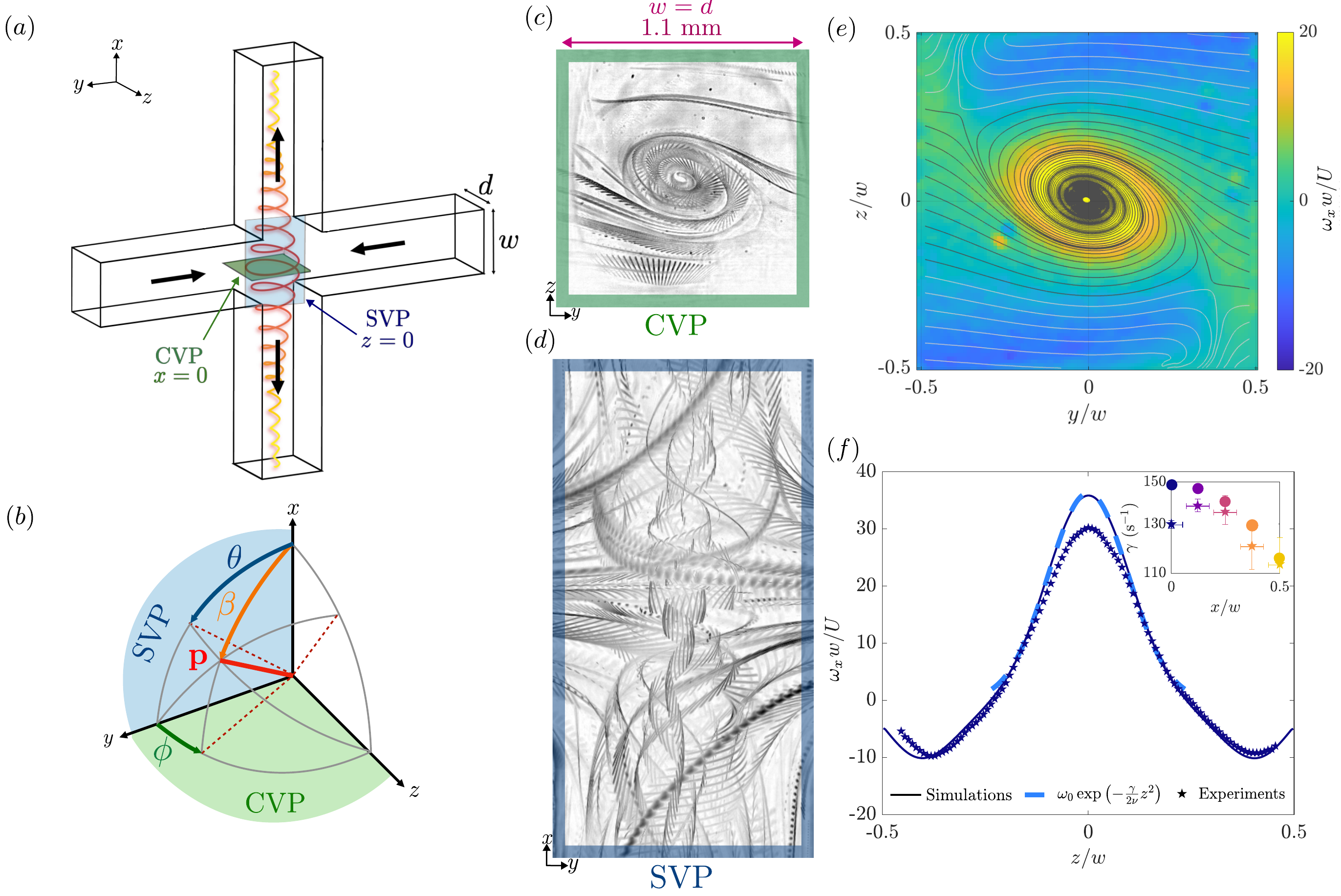}
\caption{(a) Particle flow in the 3D vortex formed in the cross-slot geometry, with the vortex stretched along the opposing outlets. 
(b) Definition of the fiber orientation vector $\mathbf{p}$ and the polar angle $\beta$ between the fiber and the vortex axis. The projected angles $\theta$ and $\phi$ are measured experimentally in the SVP and CVP, respectively. (c,d) Superposition of experimental images of rigid fibers at $\Rey=46$ in the Core Vortex Plane (CVP, $x=0$) and at $\Rey = 56$ in the Stretched Vortex Plane (SVP, $z=0$). (e) Base flow at $\Rey=56$: experimental normalized vorticity field with superimposed streamlines. Dark streamlines are drawn towards the vortex core. 
(f) Dimensionless vorticity profiles along $z/w$ at $y/w=0$ and $x/w=0$ for simulations (solid line) and PIV measurements (stars). The vortex core is fitted with the Burgers-vortex profile $\omega_x(z)w/U=\omega_0\exp(-z^2/r_\gamma^2)$ (dashed line), with $\omega_0=36$ and $r_\gamma=\sqrt{2\nu/\gamma}=150~\mu$m. The inset shows the strain rate $\gamma$ versus $x/w$ from simulations (circles) and PIV measurements (stars). Horizontal error bars indicate the uncertainty in $x$ due to the finite depth of field ($\simeq0.1w$), and vertical error bars the corresponding uncertainty in $\gamma$ estimated from the simulations.}
\label{fig:figure1}
\end{figure}

In this paper, we investigate experimentally and numerically the orientation dynamics of rigid neutrally buoyant fibers in a stationary three-dimensional vortex at intermediate Reynolds number. This flow provides a controlled analogue of a Burgers-like vortex, relevant to vortical structures encountered in turbulence, enabling a precise characterization of the fiber orientation dynamics. Initial observations show that fibers tend to rotate around the vortex while aligning with a preferential direction (see experimental stacks in figure~\ref{fig:figure1}), reminiscent of observations in turbulent flows. After introducing the experimental set-up and numerical methods in section~\ref{sec:methods}, we adopt a textbook approach and derive the fiber orientation dynamics in a Burgers vortex from Jeffery equations and then compare this theoretical prediction with a complete experimental and numerical characterization in section~\ref{sec:theory} and section~\ref{sec:exp_sim_orientation}. Finally, we discuss deviations from Jeffery’s framework arising from finite particle size and fluid inertia.
 
\section{Experimental and numerical methods} \label{sec:methods}

\subsection{Experimental set-up and base flow}

In the present study, a steady three-dimensional vortex is generated using a microfluidic cross-slot geometry (figure~\ref{fig:figure1} (a)). Two opposing inlet streams directed along the $y$-axis meet at the center of the junction, producing a stagnation point. Above a critical Reynolds number, $\Rey_c$, the base flow becomes unstable and develops into a Burgers-like vortex extending downstream into the outlet channels \citep{haward2016tricritical,burshtein2017inertioelastic}. The flow Reynolds number is defined as $\Rey = \frac{\rho_f U w}{\eta}$, where $U = Q/(wd)$ is the mean flow velocity in each inlet and outlet channel, $Q$ is the imposed flow rate, and $\rho_f$ and $\eta$ are the fluid density and dynamic viscosity, respectively. Experiments are conducted in two microfluidic devices, allowing observations in two orthogonal planes: the Core Vortex Plane (CVP, $x=0$) using a glass device, and the Stretched Vortex Plane (SVP, $z=0$) using a PDMS device. Both microchannels have a square cross-section ($w=d=1.1$~mm, aspect ratio $AR=1$). The glass microfluidic cross-slot device is fabricated by selective laser-induced etching in fused silica using a LightFab 3D printer (LightFab GmbH) \citep{meineke2016,burshtein2019}. The inlets are long ($19w$) to allow for a fully developed flow in the cross-slot area, and the outlets are designed as long as possible ($5w$) while maintaining optical access for imaging in the CVP. The PDMS device is fabricated using classical lithography techniques, with both inlets and outlets designed as long as possible ($19w$). The fibers are suspended in a $25$ wt\% glycerol-DI water solution ($\rho_f = 1059$~kg~m$^{-3}$, $\eta = 1.79$~mPa~s). 
Experimental chronophotographs are shown in figure~\ref{fig:figure1} (c,d): fibers rotate around the vortex axis in the CVP plane and exhibit "ribbon-like" dynamics in the orthogonal SVP plane. For this geometry, the instability onset occurs at $\Rey_c \simeq 42$, and all experiments are performed in the range $40 < \Rey < 80$ by varying the imposed flow rate $Q$, for which the vortex remains stationary \citep{burshtein2021periodic}. Further details regarding the experimental set-ups are provided in \citep{aulnette2025transport}.

Quantitative measurements of the base flow are carried out in the CVP using micro-particle image velocimetry ($\mu$-PIV) in the glycerol water solution wihout fibers, and are complemented by direct numerical simulations of the single-phase flow \citep{aulnette2025transport}.  The vorticity field displayed in figure~\ref{fig:figure1} (e) reveals the presence of two regions in the flow: one in which streamlines are drawn towards the vortex core (dark streamlines), as opposed to one that remains essentially unaffected by the vortical structure (light streamlines).
Within the confined central region, vorticity is highly concentrated around the stagnation point, forming a narrow column aligned with the rotation axis $x$ closely following a Gaussian distribution. This description is limited to the vortex core as outside this region, the flow is increasingly affected by the confining walls. These features are consistent with the analytical Burgers vortex solution, for which vortex stretching balances viscous diffusion to maintain a localized vorticity distribution. This agreement is further illustrated in figure~\ref{fig:figure1} (f), where the transverse vorticity profiles obtained from simulations and PIV measurements at $x=0$ are compared with the Burgers-vortex profile $\omega_x = \omega_0 \exp\left(- z^2 / r_\gamma^2\right)$, with $r_\gamma = \sqrt{2 \nu/\gamma}$ the characteristic size of the Burgers vortex and $\nu = \eta/\rho_f$ the fluid kinematic viscosity. The same analysis is then used to estimate the local strain rate $\gamma = \partial u_x/\partial x$ at different downstream positions $x/w$, as shown in the inset. The present flow exhibits a moderate downstream decrease of $\gamma$, from approximately $150$ to $115$~s$^{-1}$ over a distance of $0.5 w$ (figure~\ref{fig:figure1} (f), inset). This translates into a moderate increase of the vortex characteristic size $r_\gamma$ along $x$, from $150$~$\mu$m to $170$~$\mu$m. PIV measurements at different downstream positions show the same decrease of $\gamma$ with $x$ and remain in good overall agreement with the simulations, despite some experimental scatter. The finite depth of field results in an uncertainty of approximately $0.1w$ in the measurement position $x$ and the corresponding uncertainty in $\gamma$, estimated from its spatial variation in the simulations, are show as error bars in the inset. Although the strain rate is spatially uniform in an ideal Burgers vortex, this analytical solution remains the most relevant canonical framework for describing the structure and dynamics of the present flow.

\subsection{Rigid fibers fabrication and characterization} 
Rigid elongated fibers are fabricated in bulk from an emulsion of SU8 droplets dispersed in a glycerol-ethanol mixture \citep{alargova2006formation,li2024dynamics}. During emulsification, shear elongates the SU8 droplets, which are subsequently photo-crosslinked under UV illumination to produce stable colloidal fibers. The particles are predominantly straight, with residual curvature $\mathcal{C} = \frac{\ell_{tot}}{\ell}-1$ ($\ell_{tot}$ the total length and $\ell$ the end-to-end length) below $5\%$, and possess aspect ratios ranging from 10 to 100. This characterization is performed on images of fibers at rest. Their dimensions are controlled through the solvent composition and mixing rate. In the present study, a $70/30$ wt\% glycerol-ethanol mixture stirred at 300 rpm yields fibers of mean diameter $a \simeq 2-4~\mu$m and length $L \in [40,500]~\mu$m (see figure 1 of \citep{li2024dynamics} for size distributions). Owing to the large Young’s modulus of crosslinked SU8 ($0.9$-$7.4$~GPa), the fibers can be considered rigid under the present flow conditions. Indeed, the elastoviscous number, defined by $\bar{\eta} = 8 \pi \eta \omega_{max} L^4 / cEI$ \citep{islam2026dancing}, with $\omega_{max} \simeq 2200$ s$^{-1}$ the maximum vorticity at the core, $I = \pi a^4/64$ and $c = - \ln(e(a/L)^2)$ with $e$ the Euler number, varies between $0.017 \leq \bar{\eta} \leq 187$ (for $E = 0.9$ GPa). For comparison, simulations of flexible fibers interacting with a Burgers-like vortex by Islam et al. showed no significant deformation up to $\bar{\eta}\simeq700$. As mentioned in the previous section, the fibers are suspended in a density-matched glycerol-DI water solution, resulting in nearly neutrally buoyant particles. The suspension is sufficiently dilute that fiber-fiber interactions are negligible, allowing the fibers to be treated as isolated particles advected by the flow. 

We define the particle Reynolds number, $\Rey_p=\left(L/w \right)^2 \Rey$, using the largest particle dimension $L$ \citep{voth2017anisotropic,di2024influence,wang2025inertial} and the single phase flow Reynolds number $\Rey$. In our experiments, the particle Reynolds number is in the range $0.05~\leq~\Rey_p~\leq~12$. Another key parameter is the ratio of fiber length $L$ to vortex size $r_\gamma$, chosen at $x = 0$, which varies between 0.2 and 4 in our experiments.  
\subsection{Numerical method}
Simulations of isolated fibers of length $L$ and diameter $a$ are performed in the simulated cross-slot flow used in \cite{aulnette2025transport}. To explore similar initial conditions than the experiments, the fibers center of mass is initialized at $(y_0,z_0,x_0)/w = (-3/8,0,\epsilon)$, with $\epsilon \simeq 10^{-4}$, $\epsilon\ne0$ so that particles are not trapped in the CVP for an extended period of time, and their initial orientation is varied in the ranges $0< \beta_0<\pi/2$ and $\pi/8<\phi_0 <\pi /2$ (see sketch in figure~\ref{fig:figure1} (b)). In most experimental cases, $\beta_0 \simeq \pi/2$ (fibers enter parallel to the CVP) and $\phi \ne 0$ due to their alignment with the slightly inclined flow streamlines caused by elongation along the separatrix. The fiber diameter is set to $a = 4$ $\mu$m and the fiber length spans from $40$ to $500$ $\mu$m to match the geometrical properties of the fibers used in the experiments.
In the simulations, the particle Reynolds number is assumed to be sufficiently small for the flow around the fiber to remain in the viscous regime ($Re_p \ll 1$). Although this assumption may appear restrictive, particularly for the longest fibers considered, the results presented in the following section demonstrate that viscous effects dominate the orientation dynamics and reproduce the experimental observations accurately. The effect of fluid inertia is discussed in greater detail in the Discussion section.

The elastohydrodynamic interactions between the fiber and flow are modeled with the bead-spring model, which has been widely used, and validated, in the literature to study the motion of rigid and flexible fibers in viscous flows 
\citep{delmotte2015general,marchetti2018deformation,schoeller2021methods,slowicka2022buckling,li2024dynamics,li2026microfluidic}. Full details of the method can be found in \citep{marchetti2018deformation,li2024dynamics}, a brief overview is provided below.

The fiber is modeled as a chain of $n$ rigid spherical beads of radius $a/2$, connected by internal stretching and bending elastic forces $\mathbf{F}^E$ that enforce fiber inextensibility and rigidity over time. These forces are derived from an elastic potential $H$, whose expression is given in \citep{marchetti2018deformation,li2024dynamics}.
This model can be used to simulate both flexible and rigid fibers within the same framework by tuning the Young modulus $E$ according to the situation. In our case, we chose $E = 0.16$ GPa, which is enough to prevent fiber bending in the present flow conditions. Using the definition introduced above, the corresponding elastoviscous numbers range from $0.035$ to $355$. Despite being slightly larger than in the experiments due to the lower Young's modulus, these values remain well below those at which significant deformation was observed in \citep{islam2026dancing}.


The velocity of the fiber beads is then computed from the linear superposition principle \citep{guazzelli2011physical}: $\textbf{U}_i = \textbf{u}^{\infty}_i + \sum_{j = 1}^n \textbf{M}_{ij}\textbf{F}^E_j,$
where $\textbf{U}_i$ is the total velocity of bead $i$, $\textbf{u}^{\infty}_i$ is the background (cross-slot) flow velocity interpolated at the bead center,  and $\sum_{j = 1}^n \textbf{M}_{ij}\textbf{F}^E_j$ is the velocity induced by the nonhydrodynamics forces $\textbf{F}^E$ acting on the $n$ fiber beads. \textbf{M} is the so-called mobility matrix that contains all hydrodynamic interactions between the fiber beads, i.e.\ their velocity induced by the flow disturbances generated by the nonhydrodynamic forces. In this work, we use the Rotne-Prager-Yamakawa mobility matrix \citep{Wajnryb2013}, which is commonly used in the bead-model literature. The new position of the fiber beads \textbf{x} is then computed by integrating $d\textbf{x}/dt = \textbf{U}$.

\section{Orientation dynamics of rigid fibers in a vortex}
\subsection{Initial observations}
\begin{figure}[t]
\begin{center}
\includegraphics[width=13cm]{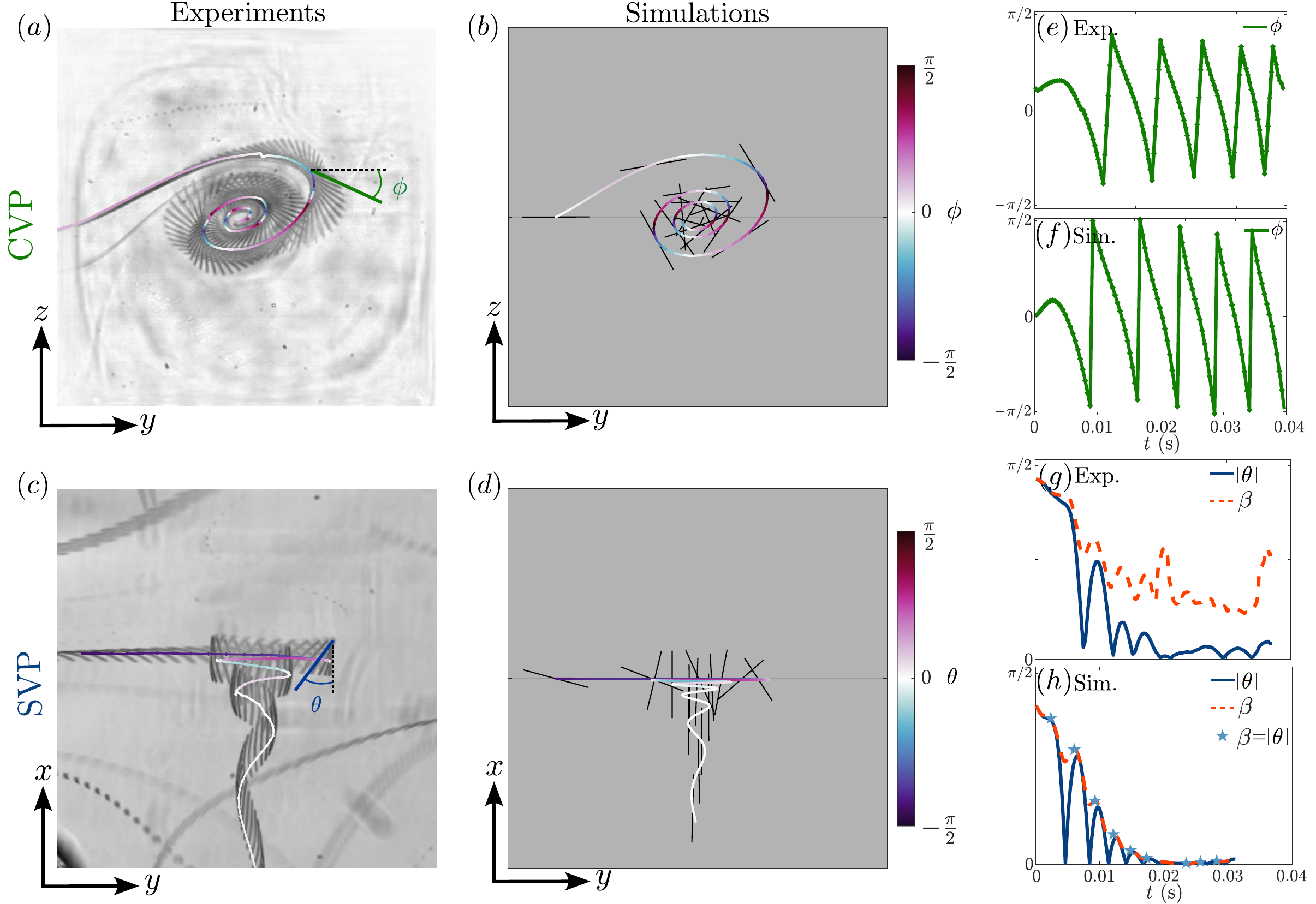}
\caption{Stacks of experimental and simulated fiber images in the cross-slot geometry. (a,c) Experimental visualization of the Core Vortex Plane (CVP), obtained with a microfluidic glass device at $\Rey = 46$ and Stretched Vortex Plane (SVP), imaged using a PDMS device at $\Rey = 56$. Overlayed are the experimental trajectories color-coded as a function of the angles $\phi$ and $\theta$ as defined in figure~\ref{fig:figure1}. (b,d) Chronophotographs of fibers overlayed with trajectories obtained from simulations are color-coded as a function of the angles $\phi$ and $\theta$ in the CVP and SVP at $\Rey = 46$ and at $\Rey = 56$, respectively. (e,g) Temporal evolutions of $\phi$ and $|\theta|$ are plotted. The dashed line represents the experimental estimation of $\beta$ from the direct measurement of $\theta$. (f,h) Temporal evolutions of $\phi$ and $|\theta|$ are plotted. The dashed line represents the time evolution of $\beta$. Blue stars (\textcolor{cyan}{$\bigstar$}) highlight the instants where $\beta = \theta$.}
\label{fig:figure2}
\end{center}
\end{figure}

In this study, we investigate the orientation dynamics of rigid elongated fibers in a steady vortex flow. Experimentally, the fiber orientation is reconstructed from complementary measurements performed in the CVP and SVP, providing access to the azimuthal angle $\phi$ and to $\theta$, the projection of the inclination angle $\beta$ onto the SVP, as defined in figure \ref{fig:figure1} (b). Figures~\ref{fig:figure2} (a,c) show experimental chronophotographs in both planes together with reconstructed trajectories color-coded by $\phi$ and $\theta$, respectively. Owing to the dilute nature of the suspension, only a few fibers are visible in each frame, such that particles may be considered isolated and individually tracked. These measurements are complemented by numerical simulations of independent fibers advected in the vortex. Representative chronophotographs are shown in figures~\ref{fig:figure2} (b,d). Both experiments and simulations reveal that fibers spiral around the vortex axis while being advected downstream, undergoing simultaneous reorientation along their trajectories.

Fibers enter the cross-slot nearly perpendicular to the vortex axis and are initially confined to the CVP ($\theta = \beta \simeq \pi/2$). As they rotate around the vortex core and migrate towards it, the angle $\phi$ oscillates between $-\pi/2$ and $\pi/2$ at each revolution with a period $\tau_\phi$ (see figures~\ref{fig:figure2} (e,f)). Simultaneously, the fibers progressively align with the vortex axis, explaining the "ribbon-like" structure observed in panel (c): $\beta$ decreases towards $0$, corresponding to alignment, as shown in figures~\ref{fig:figure2} (g,h). The experimentally accessible projected angle $\theta$ also approaches zero, but oscillates as the fiber rotates around the vortex axis. The two angles are related by $\tan\theta=\tan\beta\cos\phi$; the oscillations of $\theta$ therefore result from the projection of the rotating fiber onto the SVP, and $\beta$ and $|\theta|$ coincide when $|\cos\phi|=1$, as indicated by \textcolor{cyan}{$\bigstar$} in panel (h). Note that, in the experiments, $\beta$ was reconstructed from the apparent fiber length $L_{\rm SVP}$ using $\cos\beta=(L_{\rm SVP}/L)\cos\theta$. This reconstruction is particularly sensitive to small uncertainties in the measured fiber length and is therefore significantly noisier than the direct measurement of $\theta$, especially when the fiber aligns with the vortex axis and $L_{\rm SVP}$ approaches the total length $L$. Since $\theta$ captures the same progressive alignment towards the vortex axis, while being directly accessible experimentally, we use $\theta$ throughout the experimental analysis.

\subsection{Theory} \label{sec:theory}

To explain those dynamics, we use a simple approach combining Jeffery equations  for the orientation of a fiber in a viscous flow  and the analytical Burgers vortex velocity field. 

In Jeffery's work, anisotropic particles translate with the fluid, just as tracer particles do ($\dot{\textbf{x}} = \textbf{u}$), but their rotation depends now on their orientation with respect to the local fluid velocity gradients. We consider here prolate axisymmetric particles. Let us define \textbf{p} as the unit vector aligned with the symmetry axis of the elongated fiber (see figure~\ref{fig:figure1} (b)). The overall rotation dynamics of such a particle can be split into a spinning part (rotation around the symmetry axis) and a tumbling part (rotation of the symmetry axis) $\dot{\textbf{p}}$. In the absence of inertial and Brownian forces, Jeffery found that the symmetry axis of an axisymmetric rigid particle suspended in a simple shear flow rotates along one of an infinite one-parameter family of closed periodic orbits, known as Jeffery orbits \citep{jeffery1922}. The tumbling $\dot{\textbf{p}}$, i.e. the time evolution of the particle orientation, is given by :

\begin{equation}
\dot{\textbf{p}}
=
\boldsymbol{\Omega}\cdot\textbf{p}
+
\kappa
\left(
\textbf{E}\cdot\textbf{p}
-
(\textbf{p}\cdot\textbf{E}\cdot\textbf{p})\textbf{p}
\right),
\label{eq:Jeffery_eq}
\end{equation}

with $\boldsymbol{\Omega}$ the fluid rotation rate tensor, $\textbf{E}$ the fluid strain rate tensor and $\kappa = \frac{(L/a)^2-1}{(L/a)^2+1}$ a shape parameter depending on the aspect ratio of the length of the fiber along the symmetry axis $L$ and its diameter perpendicular to the axis $a$. Note that the second term between parentheses on the right hand side keeps $|\textbf{p}| = 1$. In the case of very elongated fibers, $L/a \gg 1$, the parameter $\kappa$ is close to 1. 

Following this definition of \textbf{p}, the unit vector  aligned with the symmetry axis of the fiber, we apply equation~\ref{eq:Jeffery_eq} to the transport of an elongated fiber in a Burgers vortex velocity field defined in cylindrical coordinates by :

\begin{align}
u_r= -\gamma r, \quad  u_\phi = \frac{\Gamma}{2\pi r}\left(1-e^{-\frac{\gamma r^2}{2\nu}}\right), \quad u_x = 2\gamma x,
\label{eq:burgers_vortex}
\end{align}

where $r = \sqrt{y^2+z^2}$ is the radial distance to the vortex center, $\gamma$ is the constant strain rate, $\Gamma$ the circulation, and $\nu$ the kinematic viscosity. The vorticity profile is given by :
\begin{equation}
    \omega_x(r) = \frac{\Gamma \gamma}{2 \pi\nu}\exp{\left(-\frac{\gamma r^2}{2 \nu}\right)}.
    \label{eq:vorticity_burgers}
\end{equation}
The Burgers vortex is characterized by the concentration of vorticity around the vortex axis in an area of characteristic size $r_\gamma = \sqrt{2\nu/\gamma}$ (i.e. the standard deviation of the Gaussian vorticity profile).

Near the vortex core, i.e. for $r \ll r_\gamma$, the azimuthal velocity and the vorticity can be approximated by :
\begin{equation}
u_\phi \approx \frac{\Gamma\gamma}{4\pi\nu} r \quad \textup{ and } \quad \omega \approx \frac{\omega_x(r=0)}{2} = \frac{\Gamma \gamma}{4 \pi \nu}.
\label{eq:approx}
\end{equation}

Each component of the particle orientation \textbf{p} provides the azimuthal and polar angles, $\phi$ and $\beta$ respectively (see figure~\ref{fig:figure1} (b)), defined as :

\begin{equation}
\textbf{p}
= (p_y,p_z,p_x) =
(\sin\beta\cos\phi,\;
 \sin\beta\sin\phi,\;
 \cos\beta).
 \label{eq:vector_p}
\end{equation}

Finally, the Jeffery equations yield :
\begin{align}
\dot p_y &= -\omega p_z - \kappa\gamma p_y
          - \kappa\gamma p_y(3p_x^2-1), \\
\dot p_z &= \omega p_y - \kappa\gamma p_z
          - \kappa\gamma p_z(3p_x^2-1), \\
\dot p_x &= 3\kappa\gamma p_x(1-p_x^2).
\end{align}


Substituting the angular parametrization (equation~\ref{eq:vector_p}) into the Jeffery equations and projecting onto the local polar direction give :
\begin{equation}
\dot{\beta}
=
- 3\kappa\gamma \sin\beta\cos\beta
=
-\frac{3}{2}\kappa\gamma\sin(2\beta).
\end{equation}

Note that this equation is independent of the flow vorticity and of the radial position of the fiber. Assuming the strain rate $\gamma$ is constant, this is integrable and we can deduce the temporal evolution of the angle $\beta$:
\begin{equation}
\tan\beta(t)
=
\tan\beta_0 \, e^{-3\kappa\gamma t}.
\label{eq:equation_beta}
\end{equation}

In a Burgers vortex, the classical closed Jeffery orbits observed in a linear shear flow, are suppressed by the axisymmetric extensional strain. Instead, the fiber orientation exponentially relaxes towards two stable fixed points $\beta = 0$ and $\beta = \pi$ corresponding to alignment with the vortex axis. Therefore, an elongated fiber will align with the vortex axis with a characteristic timescale $(3\kappa \gamma)^{-1}$ that depends on the fiber aspect ratio and the vortex strain rate.
\bigbreak
Substituting the angular parametrization (equation~\ref{eq:vector_p}) into the Jeffery equations and projecting onto one of the azimuthal directions, we show that the azimuthal angle evolves with the flow vorticity according to :
\begin{equation}
    \dot{\phi} = \omega,
    \label{eq:phi_time}
\end{equation}

Once the fiber enters the vortex core, the fluid vorticity tends towards $\omega$, so we have $\phi(t) = \phi_0 + \omega t$. The fiber therefore undergoes uniform precession about the vortex axis while simultaneously aligning with it and both processes are decoupled. This is in qualitative agreement with experimental observation described in figure~\ref{fig:figure2}. Note that, without the approximation made in equation~\ref{eq:approx}, the vorticity increases along the particle trajectory so that it spins faster as it comes close to the core. 
\subsection{Comparing dynamics in the cross-slot vortex to the idealized Burgers vortex} \label{sec:exp_sim_orientation}
\begin{figure}[t]
\centering
\includegraphics[width=10cm]{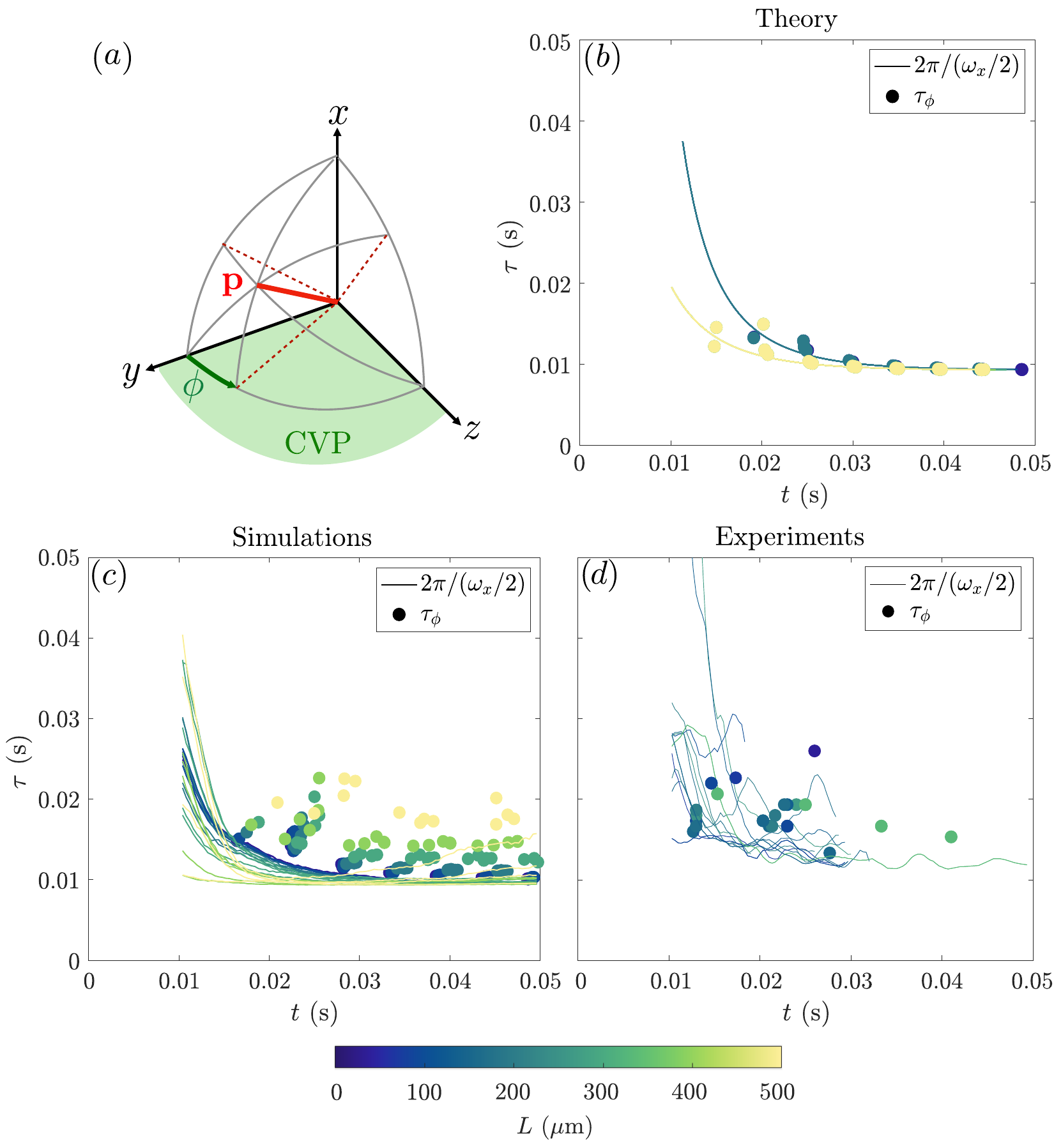}
\caption{Rotation dynamics of elongated fibers. Panel (a) defines the azimuthal angle $\phi$, measured in the CVP. Panels (b–d) show the rotation period of the fibers ($\circ$) compared with that of the fluid ($-$), as a function of time $t$ along the fiber trajectories, for (b) a Burgers vortex, with fiber dynamics predicted by Jeffery's equations, (c) the simulated three-dimensional vortex and (d) the experimental three-dimensional vortex at $\Rey=46$. In panels (b–d), the data are color-coded according to fiber length $L$, as indicated by the color bar.}
\label{fig:figure3b}
\end{figure}


\begin{figure}[t]
\centering
\includegraphics[width=10cm]{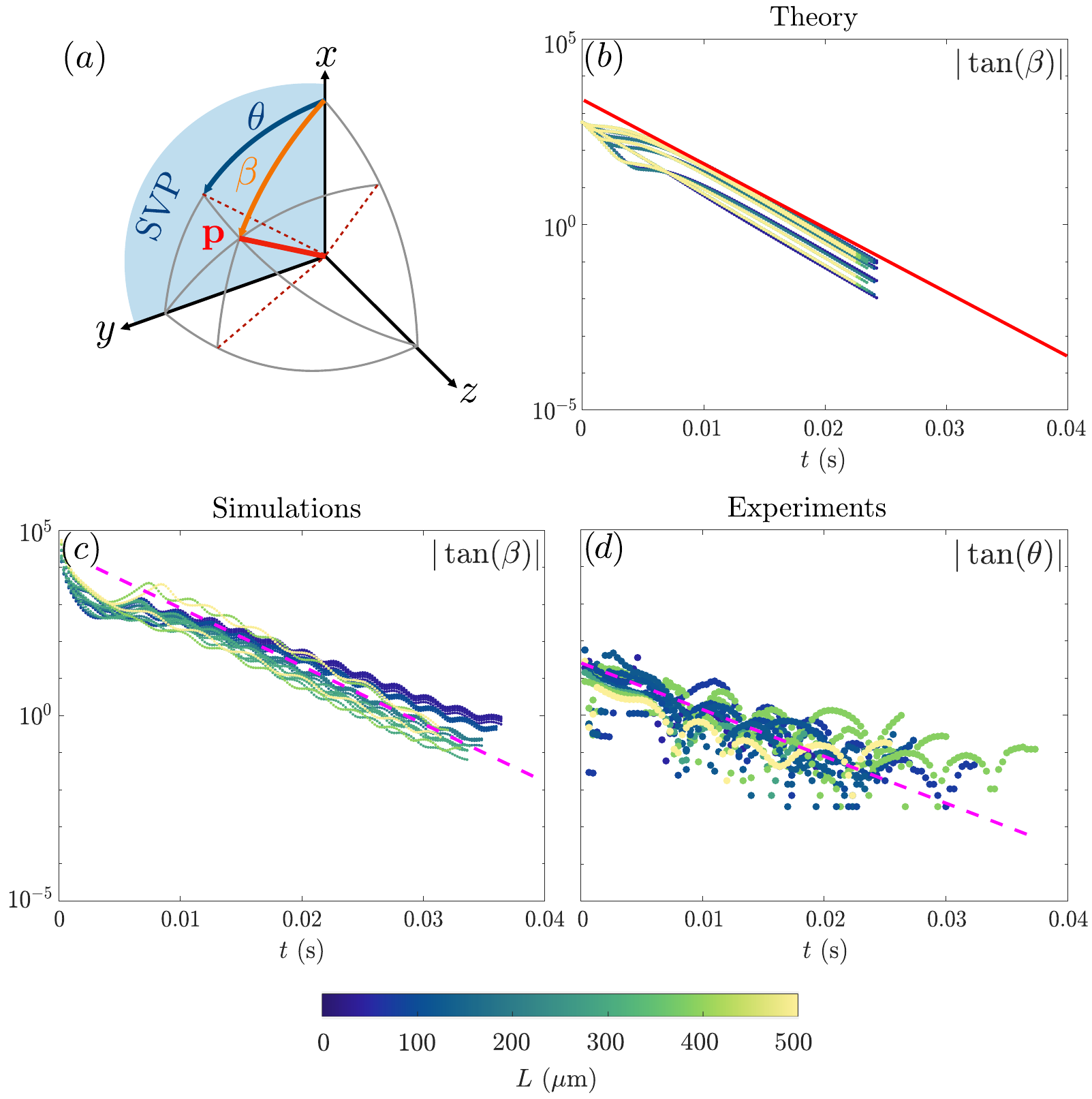}
\caption{Alignment dynamics of elongated fibers. Panel (a) defines the polar angle $\beta$ relative to the vortex axis and its projection $\theta$ onto the SVP. Panels (b–d) show the alignment dynamics for (b) a Burgers vortex, with fiber orientation predicted by Jeffery's equations, (c) the simulated three-dimensional vortex and (d) the experimental three-dimensional vortex at $\Rey=56$. The time evolution of $|\tan\beta|$ is shown in panels (b) and (c), while panel (d) shows the evolution of $|\tan\theta|$. The magenta dashed lines indicate the mean alignment rate, averaged over all trajectories, for the simulations in panel (c) and the experiments in panel (d), respectively. Data in panels (b–d) are color-coded according to fiber length $L$, as indicated by the color bar.}
\label{fig:figure3a}
\end{figure}


We now compare the dynamics predicted for a Burgers vortex with the experimental and numerical results. Figures~\ref{fig:figure3b} and~\ref{fig:figure3a} summarize the orientation dynamics for a range of fiber lengths and initial conditions by showing the temporal evolution of the fluid and fiber rotation rates (figure~\ref{fig:figure3b}) and $|\tan\beta|$ or $|\tan\theta|$ (figure~\ref{fig:figure3a}).

We first consider the azimuthal rotation dynamics in the CVP. We compare the fiber rotation period $\tau_\phi$ with the local fluid rotation period $2\pi/(\omega_x/2)$, where $\omega_x$ is obtained from the Burgers solution (equation~\ref{eq:burgers_vortex}) or from the measured and simulated velocity fields and evaluated along each fiber trajectory (figure~\ref{fig:figure3b} (b,c,d)). For the shortest fibers, $\tau_\phi$ closely follows the local fluid rotation period, in excellent agreement with equation~\ref{eq:phi_time}. As the fibers approach the vortex core, the local vorticity increases and their rotation therefore accelerates. In the simulations, the fibers eventually reach the vortex core, where the vorticity becomes nearly constant and $\tau_\phi$ approaches a constant value. In the experiments, some fibers have not yet reached this region over the observation time. Moreover, as the fibers approach the vortex core, their motion becomes increasingly three-dimensional and they progressively align with the vortex axis, making them difficult to track in the CVP. Consequently, the experimentally accessible trajectories predominantly sample the region where the local vorticity is still increasing and $\tau_\phi$ correspondingly decreases. As the fiber length increases, systematic departures from the local fluid rotation emerge, revealing finite-size effects that are discussed in section~\ref{sec:conclusion}.

We now consider the alignment dynamics in the SVP (figure~\ref{fig:figure3a}). Following a short transient associated with the fibers entering the vortex, the envelopes of $|\tan\beta|$ in theory (figure~\ref{fig:figure3a} (b)) and simulations (figure~\ref{fig:figure3a} (c)), and of $|\tan\theta|$ in the experiments (figure~\ref{fig:figure3a} (d)), decay exponentially, as predicted by Jeffery's equations. Superimposed on this overall exponential decay, oscillations are observed in both simulations and experiments. In the ideal axisymmetric Burgers vortex, $\beta$ relaxes monotonically towards zero and no oscillations are expected. The small oscillations of $\beta$ observed in the simulations result from slight deviations of the simulated base flow from an ideal Burgers vortex \citep{robinson1984stability}. In particular, the vortex core is not perfectly axisymmetric (figure~\ref{fig:figure1} (e)), causing the strain rate to vary azimuthally. As the fiber center of mass orbits around the vortex core, it therefore experiences a periodic modulation of the strain rate, resulting in the weak oscillations of $\beta$. These deviations are discussed further in appendix~\ref{sec:appendixA}. In the experiments, the directly measured quantity is instead the projected angle $\theta$. As discussed above, $\tan\theta=\tan\beta\cos\phi$, such that $|\tan\theta|$ oscillates as the fiber orientation rotates around the vortex axis, even when $\beta$ decreases monotonically. Thus, although the oscillations of $\beta$ and $\theta$ have different origins, orbital motion through the weakly non-axisymmetric flow in the former case and projection of the rotating fiber in the latter, both occur on the characteristic timescale of the local rotational flow, set by the local vorticity. The underlying decay in both cases reflects the progressive alignment of the fibers with the vortex axis. In both cases, the oscillations are superimposed on a well-defined exponential decay, allowing the underlying alignment rate to be extracted from the envelopes.

According to Jeffery's equations, the strain-induced alignment towards the vortex axis occurs on a characteristic timescale $(3\kappa\gamma)^{-1}$, providing a direct estimate of the local strain rate $\gamma$. For the highly slender fibers considered here, $0.98 \leq \kappa < 1$, so that $\kappa \approx 1$ and the relaxation rate reduces to $3\gamma$. For the ideal Burgers-vortex calculation, the strain rate is set to the maximal fitted value from the base-flow characterization, $\gamma=148~\mathrm{s}^{-1}$. Accordingly, the decay of the theoretical trajectories yields $\gamma_{\mathrm{th}}\simeq148~\mathrm{s}^{-1}$, independently of fiber length (red line in figure~\ref{fig:figure3a} (b)). Fitting the simulation results yields an average $\gamma_{\mathrm{sim}}\simeq120~\mathrm{s}^{-1}$ (magenta dashed line in figure~\ref{fig:figure3a} (c)), consistent with the range of strain rates obtained from the simulated base-flow analysis (figure~\ref{fig:figure1} (f)). Experimentally, we obtain an average $\gamma_{\mathrm{exp}}\simeq100~\mathrm{s}^{-1}$ (magenta dashed line in figure~\ref{fig:figure3a} (d)), again in good agreement with the strain rate inferred from the PIV measurements (figure~\ref{fig:figure1} (f)).

The larger theoretical value reflects the use of the maximal strain rate together with the assumption of a spatially uniform strain rate in the ideal Burgers vortex. In the cross-slot vortex, by contrast, the strain rate decreases slightly downstream, such that fibers sample progressively weaker strain along their trajectories. The somewhat lower experimental value compared with the simulations is also consistent with the PIV measurements, which yield slightly lower values of $\gamma$ than the numerical simulations, within the experimental uncertainties. Despite these variations, the three estimates are in good quantitative agreement.

The longer oscillation period observed experimentally compared with the simulations results from differences in the fiber trajectories. Simulated fibers follow tighter trajectories because their initial position is controlled near $x=0$, whereas experimental fibers enter at variable positions and typically remain farther from the vortex core. As fibers approach the core, they experience increasing vorticity and therefore faster rotational dynamics, resulting in a shorter characteristic period in the simulations. This difference is discussed further in appendix~\ref{sec:appendixB}.

The simulations reveal a slight increase of the relaxation rate with fiber length. A similar trend is observed experimentally, although it is less clearly resolved because of the larger measurement uncertainties and the limited number of observations for the longest fibers. This length dependence is discussed further in section~\ref{sec:conclusion}.

\section{Discussion and concluding remarks} \label{sec:conclusion}

\begin{figure}[t]
\centering
\includegraphics[width=13cm]{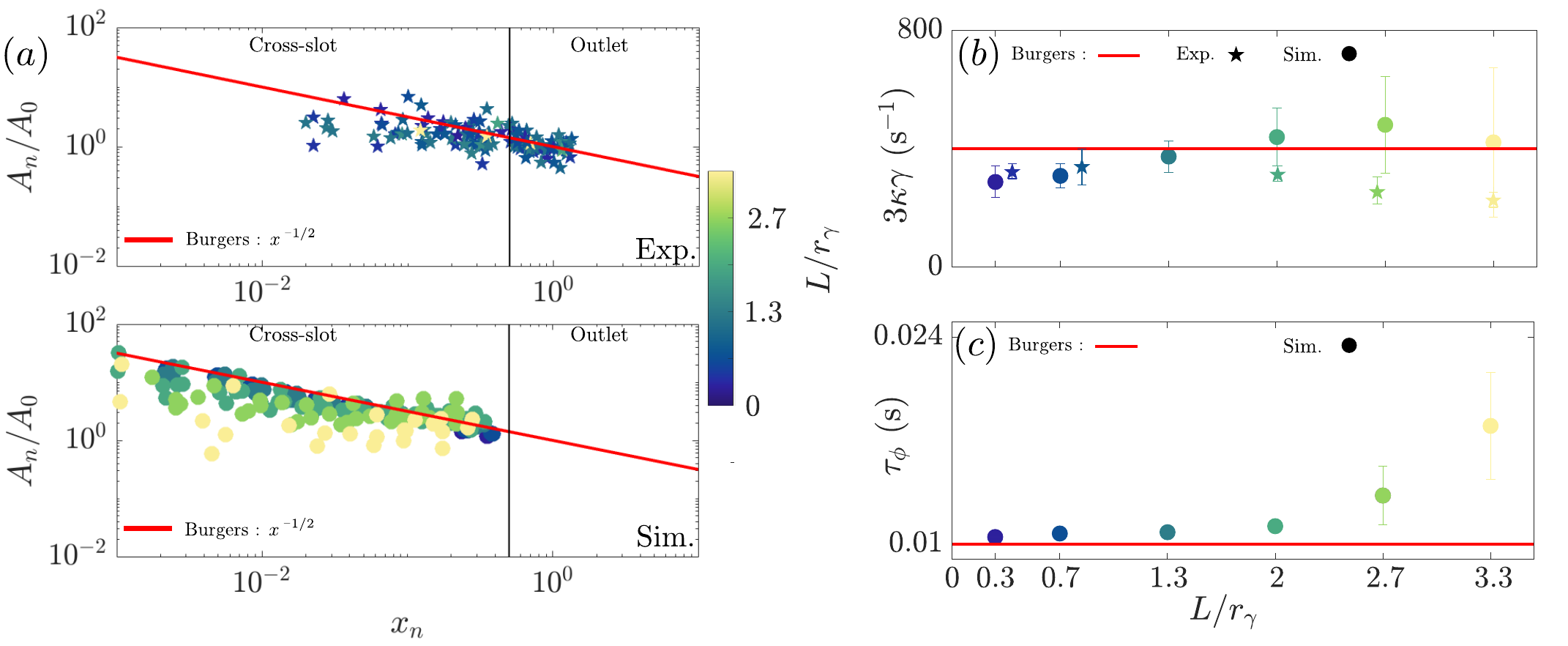}
\caption{(a) Normalized experimental and simulated amplitude of trajectories envelope in the cross-slot vortex, $A_n / A_0$ as a function of position $x_n$ in log-log scale, compared with a decaying envelope of the Burgers vortex streamlines (red solid line). The normalization coefficient $A_0$ is the fitting coefficient of each power law $A_n = A_0 x_n ^{-1/2}$. All data is color-coded with the ratio $L/r_\gamma$ comparing the fiber length $L$ to the vortex core size $r_\gamma$. (b) Exponential relaxation rate of alignment $3\kappa\gamma$ obtained from experiments (\textcolor{black}{$\bigstar$}) and simulations (\textcolor{black}{$\huge{\bullet}$}) as a function of $L/r_\gamma$. Error bars indicate the standard deviation between individual fiber measurements. For the largest fiber lengths, the experimental values are based on at most three observed fibers.  The red line represents the value $3\kappa\gamma \simeq 440$ s$^{-1}$ obtained for the Burgers vortex. (c) Average period of oscillations on $\phi$ and its standard deviation extracted from simulations (\textcolor{black}{$\huge{\bullet}$}) as a function of $L/r_\gamma$. The red line represents the value $\tau_\phi \simeq 10$ ms obtained for the Burgers vortex.}
\label{fig:figure4}
\end{figure}

We have examined the orientation dynamics of elongated fibers in a three-dimensional stationary vortex through experiments and simulations. Overall, the experimental and numerical results demonstrate that orientation dynamics of elongated fibers in the present vortex are governed by the combined action of fluid rotation and axial strain. While the azimuthal motion is controlled by the local vorticity, the strain field progressively aligns fibers with the vortex axis. Despite the finite Reynolds number, the non-ideal nature of the flow and possible finite size effects, the observed dynamics remain remarkably well captured by the canonical Burgers-vortex framework and by Jeffery’s description. 

We now discuss possible deviations from this simple picture. The Burgers-vortex description is a priori valid for tracer particles transported along streamlines. For such particles, the trajectories satisfy the power law $r = x^{-1/2}$, indicating that particles are progressively drawn towards the vortex axis as they are transported. Although the simulated fiber trajectories broadly follow this trend, larger fibers exhibit a weaker radial decay indicating that similarly to spherical particles, larger rods tend to be expelled from the vortex core (figure~\ref{fig:figure4} (a)), revealing finite-size effects even in the absence of particle inertia. In the experiments, finite-size effects may be further enhanced by inertial lift forces, which are known to promote cross-streamline migration of finite-size particles \citep{aulnette2025transport,lashgari2017inertial,tohme2021review,hur2011inertial}. However, the particle Reynolds number remains small in the present experiments, suggesting that these inertial effects are weak. Together with the smaller range of fiber lengths investigated experimentally and the measurement noise, this prevents a clear identification of these deviations.

Furthermore, Jeffery equations rely on local velocity gradients evaluated at the particle center of mass and assume that the flow is locally uniform over the particle length. As the fiber length increases, this local description may become less accurate because the fiber samples varying velocity gradients along its axis. Our simulations, which span a wider range of fiber lengths and provide higher resolution than the experiments, reveal a weak dependence of both the azimuthal rotation period and the alignment rate on $L/r_\gamma$ (figure~\ref{fig:figure4}(b,c)): longer fibers rotate slightly slower and align slightly faster than predicted by the local Jeffery description. The experimental measurements of the alignment rate show the same trend, although the values at the largest $L/r_\gamma$ are based on only a small number of fibers. The separate experimental data set used to determine $\tau_\phi$ does not provide sufficient statistics over the different fiber lengths to reliably resolve its dependence on $L/r_\gamma$. The faster alignment of longer fibers is consistent with the observations of \citep{islam2026dancing}. Bin Islam et al. additionally showed that increasing fiber flexibility further accelerates alignment with the vortex axis.
 
In simple shear flows, the existence of an infinite number of marginally stable Jeffery orbits makes the orientation dynamics highly sensitive to weak perturbations arising from particle shape, inertia or other effects, which may drive particles across Jeffery orbits \citep{zottl2019dynamics,einarsson2015effect,di2024influence}. In contrast, the stretched vortex considered here possesses a stable orientational attractor. The excellent agreement between theory, experiments and simulations indicates that the orientation dynamics remain remarkably robust despite finite particle size, finite Reynolds number, and deviations of the flow from an ideal Burgers vortex. Jeffery equations therefore provide an accurate description of the orientation dynamics over the range of particle Reynolds numbers and fiber lengths investigated here, suggesting that vortex stretching and strain stabilize the orientational dynamics. More generally, this robustness suggests that Jeffery's description may remain applicable in more complex vortical flows such as turbulence \citep{wang2024localization}.

Finally, a natural extension of this study would be to investigate the behavior of flexible fibers, particularly how their shape responds to the varying vorticity and strain encountered in stretched vortices \citep{Marchioli2026,islam2026dancing,ibarra2026trapping}. Such work would extend the present framework to natural and synthetic filaments, where flexibility may substantially alter both alignment and migration dynamics.

\section*{ACKNOWLEDGMENTS}
We thank Arash Belizad Banaei and Luca Brandt for their simulations data of the single phase base flow. We thank Noa Burshtein for fruitful discussions, as well as Simon J. Haward and Amy Q. Shen. AL and MA acknowledge support by the Institut Carnot IPGG (Institut Pierre-Gilles de Gennes), through the Carnot program funded by the Agence Nationale de la Recherche (ANR) and the Ministère de l’Enseignement Supérieur et de la Recherche (MESRI) and the Major Research Program of PSL Research University "IPGG" launched by PSL Research University and implemented by ANR with the references ANR-10-IDEX-0001. This work benefited from the technical contribution of the joint service unit IPGG Technological Platform CNRS UAR 3750. BD acknowledges support from the French National Research Agency (ANR), under awards ANR-20-CE30-0006 and ANR-25-CE30-4710.

\section*{Declaration of Interests}
The authors report no conflict of interest.

\appendix
\section{Orientation dynamics in an elliptic stretched vortex} \label{sec:appendixA}

Figure~\ref{fig:figure3a} shows that, in both the experiments and numerical simulations of the cross-slot vortex, the angle $\beta$ decays in close agreement with the theoretical predictions obtained from the analytical solution of Jeffery equations derived in the main text for the core region of an axisymmetric Burgers vortex. However, $\beta$ also exhibits small-amplitude temporal oscillations.
Although secondary to the overall decay dynamics, this behavior is not captured by the analytical solution and therefore constitutes a deviation from the theoretical prediction.

As shown in figure~\ref{fig:figure1}, the vortex core is not perfectly axisymmetric. The purpose of this section is therefore to investigate whether this lack of axisymmetry can account for the observed oscillations of $\beta$.
To investigate the effect of asymmetry, we consider the stretched elliptic vortex flow introduced by Robinson and Saffman \citep{robinson1984stability}, which extends the Burgers vortex to asymmetric configurations and qualitatively resembles the core of a cross-slot flow.

\subsection{Stretched elliptic vortex flow}

\begin{figure}[t]
\centering
\includegraphics[width=\textwidth]{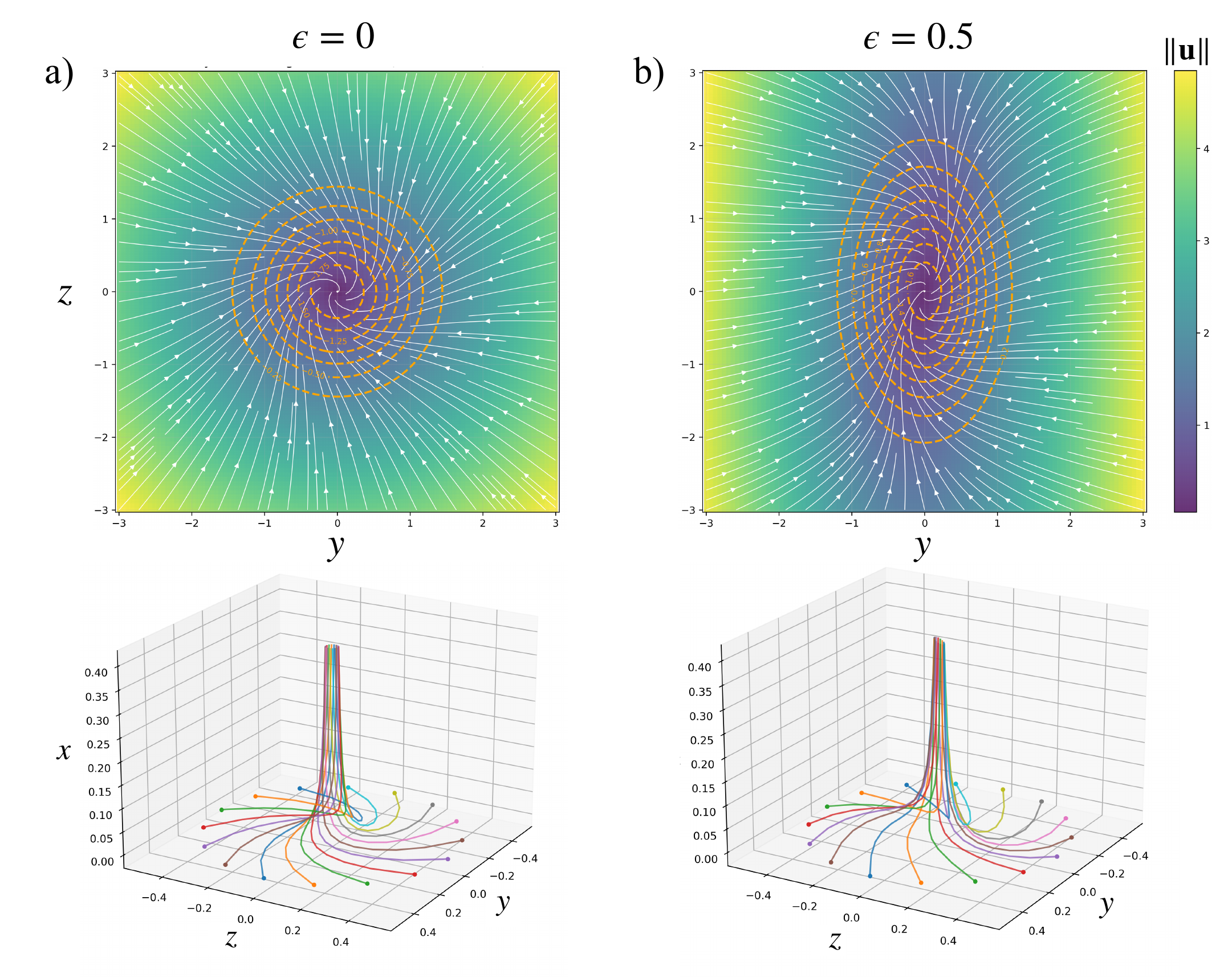}
\caption{Top: velocity field and vorticity contours (orange dashed lines) for (a) an axisymmetric stretched vortex ($\epsilon=0$) and (b) an asymmetric stretched vortex ($\epsilon=0.5$). Bottom: representative tracer trajectories.}
\label{fig:figureS1}
\end{figure}
 The steady stretched elliptic vortex flow can be decomposed into a known irrotational component, corresponding to a pure straining flow $\hat{\bu}_S$, and a vortical component $\hat{\bu}_V$ rotating about the $x$-axis:
\eqn{
\hat{\bu} &= \hat{\bu}_S + \hat{\bu}_V, \\
&= \left(\begin{array}{c}
(\alpha+\beta)\hat{x}\\
-\alpha \hat{y}\\
-\beta \hat{z}\\
\end{array}\right) + \left(\begin{array}{c}
0\\
\hat{u}(\hat{y},\hat{z})\\
\hat{v}(\hat{y},\hat{z})\\
\end{array}\right)
}
where $\alpha$ and $\beta$ denote strain rates, and hats indicate dimensional quantities.

In this flow, the only nonzero component of the vorticity is $\hat{\omega} = \hat{v}_{\hat{y}}-\hat{u}_{\hat{z}}$.

Using the characteristic time  scale $t_c = 2/(\alpha + \beta)$ and length scale $l_c = \sqrt{2\nu/(\alpha + \beta)}$, the dimensionless velocity field becomes 
\eqn{
\bu & = \left(\begin{array}{c}
2x\\
-(1+\epsilon) y\\
-(1-\epsilon) z\\
\end{array}\right) + \left(\begin{array}{c}
0\\
u(y,z)\\
v(y,z)\\
\end{array}\right),
}
where the  asymmetry parameter $\epsilon = \frac{\alpha-\beta}{\alpha+\beta}$ quantifies the anisotropy of the straining flow in the $(y,z)$ plane. The circulation-based Reynolds number is defined as Re$_{\Gamma}=\frac{\Gamma}{2\pi \nu} = \frac{\omega_0}{\gamma}$, where $\Gamma$ is the vortex circulation, and $\omega_0$ the core vorticity.

In the limit $\mathrm{Re}_{\Gamma}\rightarrow0$ (core vorticity $\ll$ strain), the steady dimensionless Helmholtz equation for the vorticity takes the form
\eqn{
\omega_{yy} + \omega_{zz}  +  (1+\epsilon)y \omega_y + (1-\epsilon)y \omega_z + 2\omega  = 0,
}
whose solution is given by \citep{robinson1984stability} 
\eqn{
\omega = \sqrt{1-\epsilon^2}\exp\left[-(1+\epsilon)\frac{y^2}{2}-(1-\epsilon)\frac{z^2}{2}\right].
}
When $\epsilon\neq0$ (i.e. $\alpha\neq\beta$), the vorticity follows an anisotropic Gaussian distribution, in contrast to the axisymmetric distribution obtained for $\epsilon=0$ and commonly associated with Burgers vortices.

Once the vorticity field is known, the vortical velocity field can be recovered from its streamfunction $\Psi$, which satisfies
\eqn{
\nabla^2\Psi &= \omega,\\
u &= -\Psi_z,\;\; v = \Psi_y.
}
This yields 
\eqn{
u(y, z) &= z \int_0^{\infty}
  \frac{\exp\!\left(-\frac{y^2}{B_+} - \frac{z^2}{B_-}\right)}
  {\sqrt{B_+}\, B_-^{3/2}} \, ds, \label{eq:usol} \\
v(y, z) &= -y \int_0^{\infty}
  \frac{\exp\!\left(-\frac{y^2}{B_+} - \frac{z^2}{B_-}\right)}
  {B_+^{3/2}\, \sqrt{B_-}} \, ds, \label{eq:vsol} \\
\text{where }B_{\pm}(s) &= s + \frac{1}{1 \pm \varepsilon} \nonumber.
}
Although these expressions do not admit a closed-form analytical representation, they can be evaluated efficiently using standard Gauss--Legendre quadrature.

Figure \ref{fig:figureS1} shows the flow field of the elliptic vortex with the vorticity contours in the plane $x=0$ (CVP), and representative tracer trajectories in the vortex.

\subsection{Orientation dynamics}
Using the velocity field defined by Eqs.\ \eqref{eq:usol}-\eqref{eq:vsol}, 
the evolution of $\beta$ is obtained by numerically integrating Jeffery equations (Eq.~3.1 of the main text) with a fourth-order Runge--Kutta scheme.

\begin{figure}[t]
\centering
\includegraphics[width=\textwidth]{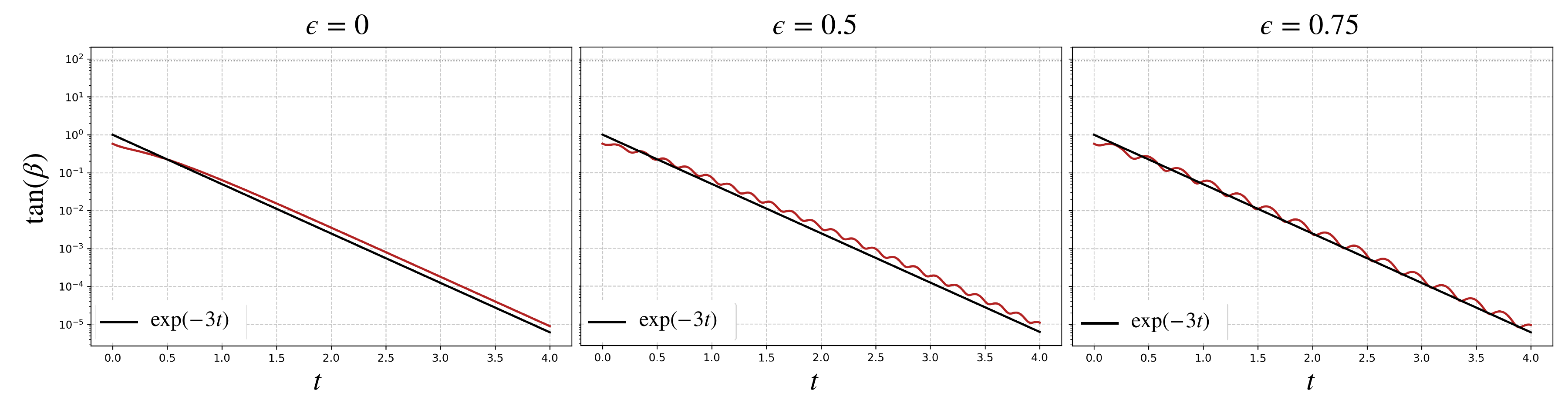}
\caption{Time evolution of $\tan\beta$ for three values of the asymmetry parameter, $\epsilon=0$, $0.5$, and $0.75$. Red lines denote numerical solutions of Jeffery  equations, and black lines indicate the theoretical exponential decay predicted for an axisymmetric Burgers vortex.}
\label{fig:figureS2}
\end{figure}

Figure \ref{fig:figureS2} shows the evolution of $\tan\beta$ for three values of the asymmetry parameter, $\epsilon=0$, $0.5$, and $0.75$. As soon as axisymmetry is broken, oscillations appear around the exponential decay predicted for the axisymmetric vortex. These oscillations closely resemble those observed in the cross-slot experiments and simulations (see Fig.\ 3 of the main text), suggesting that vortex asymmetry is a plausible mechanism underlying the observed dynamics.

\section{Oscillation period} \label{sec:appendixB}
\begin{figure}[h]
\begin{center}
\includegraphics[width=12cm]{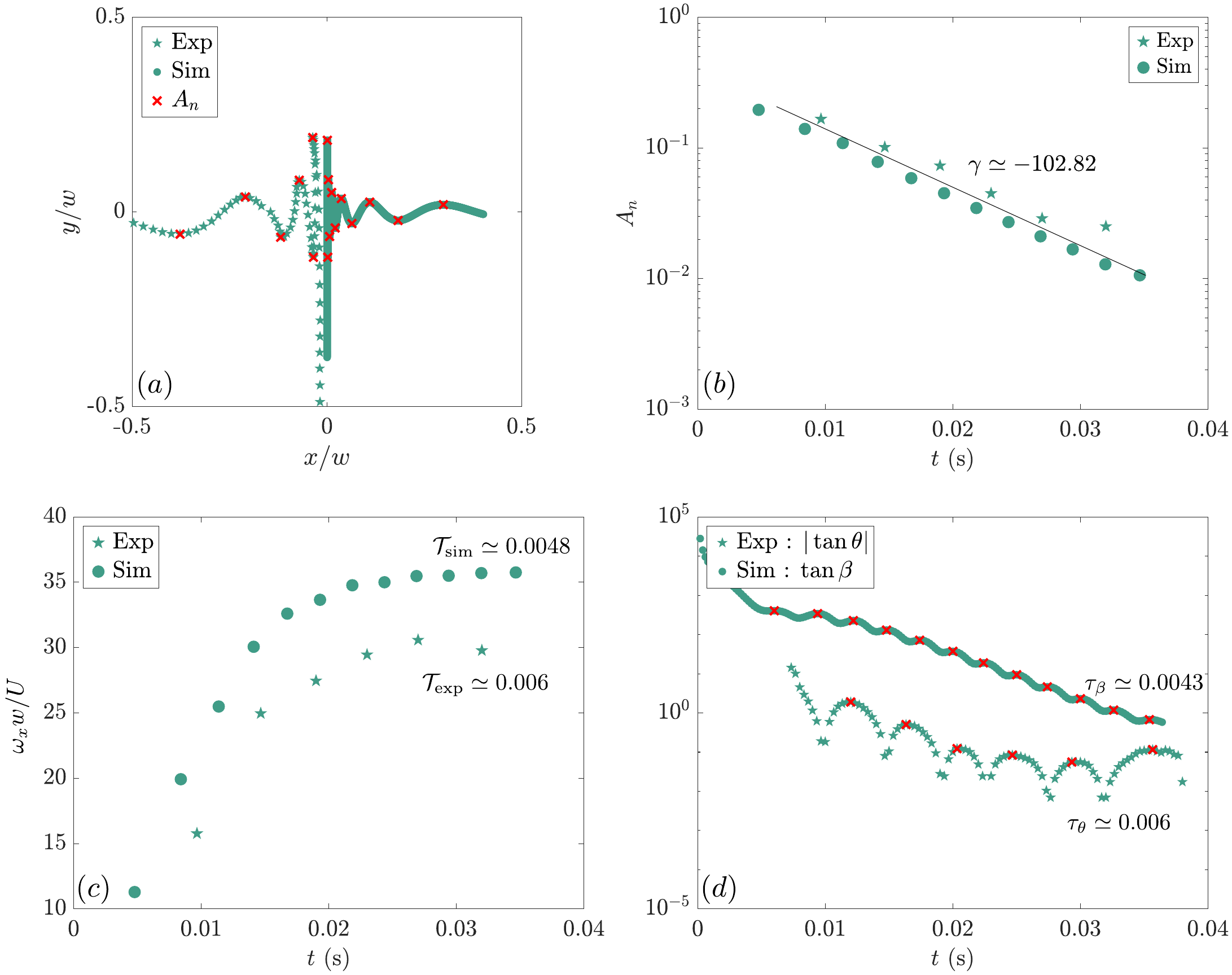}
\caption{(a) Experimental and simulated trajectories. Red crosses indicate the amplitude of the envelope $A_n$ at each turn around the vortex axis. (b) Trajectories envelope $A_n$ as a function of time in log-lin scale. The slope gives a measurement of the strain rate $\gamma$. (c) Local fluid vorticity along the fibers trajectories. Asymptotic rotation periods $\mathcal{T} = 2 \pi / (\omega_x/2)$ are written for each signal. (d) $\tan \beta$ and $|\tan \theta|$ as a function of time respectively for the simulated and experimental fiber trajectories. Oscillation periods $\tau_\beta$ and $\tau_\theta$ are measured between each maxima indicated by ($\times$).}
\label{fig:figure_oscillations}
\end{center}
\end{figure}
 As the fiber is transported into the vortex, its orientation dynamics result from two distinct mechanisms, as illustrated in figure~\ref{fig:figure2}. The first is the rotation induced by the local vorticity, which governs the azimuthal angle $\phi$ observed in the CVP. Its evolution is directly related to the local vorticity through $\dot{\phi}=\omega_x/2$. The local vorticity also sets the orbital motion of the fiber around the vortex axis, and hence its characteristic rotation period
$\mathcal{T} = 2 \pi/(\omega_x/2)$.
The second mechanism is the progressive alignment of the fiber with the vortex axis due to the extensional component of the flow. In a perfectly axisymmetric Burgers vortex, the three-dimensional angle $\beta$ therefore relaxes monotonically towards zero, without oscillations. In the experiments, however, $\beta$ is not measured directly; instead, we measure the angle $\theta$ of the fiber projected onto the observation plane. The two angles are related through $\tan\theta=\tan\beta\cos\phi$. Consequently, even when $\beta$ decreases monotonically, $\theta$ oscillates as the fiber rotates around the vortex axis, with a period set by the same rotational dynamics as $\phi$.
This picture is slightly modified by the weak ellipticity of the vortex in both experiments and simulations, which leads to the small oscillations observed in $\beta$ (cf. Appendix \ref{sec:appendixA}). While the oscillations of $\theta$ and $\beta$ thus have different origins, they both follow the orbital motion of the fiber around the vortex axis and therefore exhibit the same characteristic period, set by the local vorticity.

Figure~\ref{fig:figure3a} shows differences in oscillation period between $\beta$ and $\theta$. We now demonstrate that the observed difference stems from variations between trajectories owing to their different radial positions. Figure~\ref{fig:figure_oscillations} compares one experimental and one numerical trajectory. Panel (a) shows their projections in the SVP. The simulated trajectory forms tighter loops around the vortex core than the experimental one. As shown in \citep{aulnette2025transport}, the number of loops completed around the core depends sensitively on the initial position within the cross-slot, particularly along the vortex axis $x$. This position is readily controlled in the simulations, where fibers are initialized as close as possible to $x=0$, resulting in trajectories that remain close to the core. Experimentally, the initial position cannot be controlled and fibers rarely enter the cross-slot at $x=0$, leading to trajectories farther from the core and consequently fewer, wider loops. Panel (b) shows the corresponding radial envelopes. Assuming the flow is well described by a Burgers vortex, the radial position is expected to decay as $y(t)=\exp(-\gamma t)$, where $\gamma$ is the local strain rate. The measured envelopes are in excellent agreement with this prediction, yielding the same value of $\gamma$ for the experimental and numerical trajectories and further confirming the close correspondence between the two flows. Because the vorticity varies strongly with radial position, the local rotation rate also differs between the two trajectories. As shown in panel (c), the vorticity increases as the fibers approach the vortex core before reaching a plateau corresponding to their asymptotic rotation period $\mathcal{T} = 2 \pi / (\omega_x /2)$. The experimental fiber follows a wider trajectory and therefore samples lower vorticity $\omega_x$ than the simulated fiber, resulting in a longer rotation period $\mathcal{T}$. Finally, panel (d) shows $\tan\beta$ for the simulation and $\tan\theta$ for the experiment. Both experimental and numerical signals exhibit oscillations of period $\tau_\theta$ and $\tau_\beta$, respectively. Although the oscillation period differ between the two trajectories, they are consistent with the rotation periods $\mathcal{T}$ inferred independently from the local vorticity in panel (c): $\mathcal{T}_{\rm sim} \approx \tau_{\beta}$ and $\mathcal{T}_{\rm exp} \approx \tau_{\theta}$.


\bibliographystyle{jfm}

\end{document}